# Observation of superconductivity in pressurized Weyl semimetal candidate TaIrTe$_4$

Shu Cai[1,3], Eve Emmanouilidou[2], Jing Guo[1], Xiaodong Li[4], Yanchuan Li[4], Ke Yang[5], Aiguo Li[5], Qi Wu[1], Ni Ni[2], Liling Sun[1,3,6]†

[1]Institute of Physics and Beijing National Laboratory for Condensed Matter Physics, Chinese Academy of Sciences, Beijing 100190, China
[2]Department of Physics and Astronomy and California Nano Systems Institute, University of California, Los Angeles, California 90095, USA
[3]University of Chinese Academy of Sciences, Beijing 100190, China
[4]Institute of High Energy Physics, Chinese Academy of Sciences, Beijing 100049, China
[5]Shanghai Synchrotron Radiation Facilities, Shanghai Institute of Applied Physics, Chinese Academy of Sciences, Shanghai 201204, China
[6]Songshan Lake Materials Laboratory, Dongguan, Guangdong 523808, China

Here we report the observation of superconductivity in pressurized type-II Weyl semimetal (WSM) candidate TaIrTe$_4$ by means of complementary high-pressure transport and synchrotron X-ray diffraction measurements. We find that TaIrTe$_4$ shows superconductivity with transition temperature ($T_C$) of 0.57 K at the pressure of ~23.8 GPa. Then, the $T_C$ value increases with pressure and reaches ~2.1 K at 65.7 GPa. *In situ* high-pressure Hall coefficient ($R_H$) measurements at low temperatures demonstrate that the positive $R_H$ increases with pressure until the critical pressure of the superconducting transition is reached, but starts to decrease upon further increasing pressure. Above the critical pressure, the positive magnetoresistance effect disappears simultaneously. Our high pressure X-ray diffraction measurements reveal that, at around the critical pressure the lattice of the TaIrTe$_4$ sample is distorted by the application of pressure and its volume is reduced by ~19.2%, the value of which is predicted to result in the change of the electronic structure significantly. We propose that the pressure-induced distortion in TaIrTe$_4$ is responsible for the change of topology of Fermi surface and such a change favors the emergence of superconductivity. Our results clearly demonstrate the correlation among the lattice distortion, topological physics and superconductivity in the WSM.

**Introduction**

Weyl semimetal (WSM) is a unique material that hosts Weyl fermions as quasiparticle excitations and an exotic surface-state band structure containing topological Fermi arcs[1-4], which extends the classification of topological phases[5]. Recently, two types of WSMs with distinct band structures have been discovered in the real materials[6-19]. The family of MX (M=Ta, Nb and X=As and P) has been predicted theoretically and identified experimentally as type-I WSMs[6-13], featuring the shrinking of the bulk Fermi surface to a point at the Weyl node[2,8,10], while the family of $MTe_2$ (M = Mo and W)[14-19] and $MXTe_4$ (M = Ta and Nb; X= Ir, Rh)[20-24] has been proposed and experimentally confirmed as type-II WSMs, with tilted Weyl cones appearing at the boundaries between electron and hole pockets by breaking Lorentz invariance[14,25]. The electronic structure of WSMs gives rise to fascinating phenomena in transport properties, including a chiral anomaly in the presence of parallel electric and magnetic fields, positive magnetoresistance, a novel anomalous Hall response, surface-state quantum oscillations and exotic superconductivity[26-33], providing a research platform to promote the potential applications in spintronics or new types of topological qubits[34,35].

Intensive efforts have been made to search for Weyl superconductors in all WSMs, however, applying chemical doping for the WSMs produces a limited result[31]. Pressure is a clean and effective way to realize the tuning of interactions among multiple degrees of freedom in solids without introducing chemical complexity, and has thus been successfully adopted in the studies of many materials[36-47]. Compelling examples of pressure-enhanced superconducting transition temperature ($T_C$) in known superconductors have been observed in the families of copper oxide (cuprate) and iron-based superconductors. For instance, the $T_C$ of the mercury bearing cuprates with a value of 134 K at ambient pressure is increased to 164 K at ~ 30 GPa, which holds the highest $T_C$ among all the copper oxide superconductors[48-50]. The pressure-induced superconducting transition after suppression of large magnetoresistance in compressed WTe$_2$[51,52] and MoTe$_2$[29] are also worth noting. As a ternary variant of WTe$_2$, TaIrTe$_4$ crystallizes in an orthorhombic unit cell and can be viewed as a cell-doubling derivative. Thus, it is proposed to be a candidate of type-II WSM [20,21,22,53], hosting a combination of twelve Weyl points and two Dirac nodal rings in the Brillouin zone[53-57], and displaying a non-saturating magnetoresistance effect[53,58]. However, there is no report on its properties under high pressure conditions. Here we demonstrate experimentally the finding of superconducting transition and the corresponding changes of transport and structural properties in pressurized TaIrTe$_4$ by the complementary measurements of high-pressure resistance, alternating current (*ac*) susceptibility, magnetoresistance, Hall coefficient and synchrotron X-ray diffraction.

**Results**

We first performed temperature-dependent resistance measurements on the single crystals of TaIrTe$_4$ in a diamond anvil cell (DAC) with large culet size of the anvils. As shown in Fig.1a, the resistance at high temperature decreases with increasing pressure over the entire temperature range. No superconductivity is observed at pressure below 19.2 GPa, a maximum pressure of the anvils employed. To reveal higher pressure behavior of the TaIrTe$_4$ sample, we loaded the second sample in a DAC with small culet size of anvils and conducted higher pressure resistance measurement up to 38.6 GPa. As shown in Fig.1b, the plots of temperature versus resistance display the same behavior at pressure below ~23.8 GPa. Zooming in the low temperature resistance, we find a resistance drop starting at 23.8 GPa (Fig.1c) which becomes pronounced upon further compression (Fig.1d). At 33.6 GPa, zero resistance is observed, an evidence of superconducting transition. The zero resistance behavior is also observed in the measurements on the third sample obtained from different batches (inset of Fig.1d).

To characterize whether the pressure-induced resistance drop is associated with a superconducting transition, we applied magnetic fields on the compressed TaIrTe$_4$ subjected to 45 GPa. As shown in Fig.2a, the resistance drop temperature shifts to lower temperature upon increasing magnetic field and completely vanishes at 0.5 T. To further support that the resistance drops observed in pressurized TaIrTe$_4$ are related to a superconducting transition, we performed high-pressure *ac* susceptibility measurements in a cryostat whose lowest temperature is ~1.5 K. As shown in Fig.2b, visible diamagnetism is observed at ~ 1.60 K and 1.62 K for TaIrTe$_4$ pressurized at 44.5 GPa and 52.4 GPa, respectively. No superconducting transition is observed for the sample subjected to 40.4 GPa because its $T_C$ value (1.3 K) is lower than 1.5 K. These results indicate that the observed pressure-induced resistance drop truly originates from a superconducting transition. We estimated the upper critical magnetic field ($H_{c2}$) for the superconducting phase of TaIrTe$_4$ by using the Werthamer-Helfand-Hohenberg (WHH) formula[59]: $H_{c2}^{WHH}(0)=-0.693T_C(dH_{C2}/dT)_{T=Tc}$. The plots of $H_{C2}$ versus $T_C$ obtained at different pressures are present in the inset of Fig.2a. The estimated values of the upper critical fields of the TaIrTe$_4$ sample at zero temperature are ~ 0.83 T at 45 GPa.

Structural stability is one of the key issues for understanding the superconductivity found in the pressure range of our experiments. We thus performed high pressure X-ray diffraction measurements on the TaIrTe$_4$ sample up to 66.8 GPa. As is shown in Fig.3a, TaIrTe$_4$ crystallizes in an orthorhombic lattice with *a* = 3.80 Å, *b* = 12.47 Å, and *c* = 13.24 Å at ambient pressure[53, 60, 61]. The XRD patterns collected at different pressures are displayed in Fig. 2b. It is found that all peaks observed at pressure below 23.3 GPa can be well indexed in TaIrTe$_4$'s known ambient-pressure phase, *i.e.* the orthorhombic phase in the *Pmn2$_1$* space group. The lattice parameters and volume as a function of pressure up to 23.3 GPa are shown in Fig.3c and 3d. However, we found that some peaks in the diffraction patterns collected at pressure higher than 26.9 GPa slightly shift to small 2θ angle while no new peaks are observed, implying that the lattice distortion occurs between 23.3 GPa and 26.9 GPa. To illustrate the distortion more clearly, we extracted the pressure dependence of *d*-spacing value for different crystallographic planes (Fig.3e). It is seen that the *d* value of the (002) plane displays an apparent negative contraction starting at ~ 23.3 GPa where the superconductivity emerges. The changes of the *d*-spacing are also found in other crystallographic planes, such as (121), (041), (123) and (142), as seen in Fig.3e. These results raise the possibility that the pressure-induced lattice distortion play a crucial role for the development superconductivity in the WSM candidate TaIrTe$_4$.

**Discussions**

We summarize our high pressure experimental results obtained from measurements on TaIrTe$_4$ in the pressure-$T_C$ phase diagrams (Fig.4a). Two distinct ground states can be seen in the diagrams: the WSM state and the superconducting (SC) state. Superconductivity is not observed in the non-distorted orthorhombic (OR) ambient pressure phase below 23.8 GPa (Fig.3e and Fig.4a), but emerges in a pressure-induced distorted phase with $T_C$ about 0.57 K at 23.8 GPa. $T_C$ continuously increases with further compression and reached 2.1 K at 65.7 GPa, the maximum pressure of this study. TaIrTe$_4$ exhibits a strong anisotropy in transport properties and non-saturating magnetoresistance effect at ambient pressure[53,62-65], similar to what has been seen in WTe$_2$[58,66-70], thus it is of great interest to clarify the Hall coefficient ($R_H$) and magnetoresistance effect before and after superconducting transition because these quantities can reflect the effect of pressure on the electronic structure. Building on these ideas, we performed high-pressure Hall resistance and magnetoresistance measurements on the TaIrTe$_4$ sample by sweeping the magnetic field perpendicular to the *ab* plane up to 7 T at 10 K and different pressures, as shown in Fig.4b and Fig.S2 in SI. At ambient pressure, $R_H$ displays a positive sign at 10 K, implying that hole-carriers are dominant. Upon compression, $R_H$ increases with increasing pressure, the trend is reversed at ~22 GPa. This implies that the pressure-induced lattice distortion changes the topology of the Fermi surface, which in turn alters the population of electron carriers. Such a change seems to be in favor of developing superconductivity in TaIrTe$_4$.

A common feature of type-II WSMs is that they have a large, positive magnetoresistance[58,71,72]. Early high-pressure studies on WTe$_2$ found that superconductivity appears as the positive magnetoresistance effect is suppressed completely[51]. To understand the superconductivity in pressurized TaIrTe4, we performed high-pressure magnetoresistance measurements on our sample. As shown in Fig.4d and Fig.S3 in the SI, the ambient-pressure TaIrTe$_4$ also shows a positive magnetoresistance effect (*MR*%=24), where *MR* is defined as [(*R*(*7T*)-*R*(*0T*)/*R*(*0T*)]×100%. Upon increasing pressure, the *MR%* value decreases with elevating pressure and becomes zero at ~25.3 GPa, where the superconductivity is observed. Our results provide important evidence that the superconducting state competes with the positive MR state, *i.e.* no superconductivity develops as long as the positive magnetoresistance effect prevails. It is known that the positive magnetoresistance effect in type-II WSMs is associated with the populations of the hole and the electronic carriers[66,68,70]. Out high pressure results provide fresh information to support this scenario.

Recent theoretical calculations on TaIrTe$_4$ found that the topological band structure can be dramatically degenerated by volume change[53]. As the volume is reduced by ~ 15%, Weyl 2 points disappear and nodal lines expanse remarkably[53].

Motivated by these calculated results, we estimated the volume reduction ($\Delta V$) at ~23.3 GPa ($\Delta V = [V(23.3\ \text{GPa}) - V_0]/V_0$, where $V_0$ is the unit cell volume under ambient pressure). It is found that $\Delta V$ at ~23.3 GPa is about 19.2% (greater than 15%). To further reveal the main contribution of the lattice parameter ($a$, $b$ or $c$) to the volume shrinkage ($\Delta V$) at 23.3 GPa, we compute the corresponding $\Delta a/a$, $\Delta b/b$ and $\Delta c/c$ and find that at 23.3 GPa $\Delta a/a=4.0\%$, $\Delta b/b=5.3\%$ and $\Delta c/c=10.2\%$. These results demonstrate that the substantial reduction in the $c$ direction contributes remarkably to the degeneration of the band structure. As a result, the superconductivity found in WSM candidate TaIrTe$_4$ seems to stem from a pressure-induced electronic structure change which is driven by the lattice distortion. Clearly, the microscopic interactions under high pressure call for further experiments. Moreover, determination on whether the high-pressure distorted phase is still in an orthorhombic form is crucial because this is related to the key issue of that whether the WSM candidate TaIrTe$_4$ is a Weyl superconductor or not under pressure.

In conclusion, we are the first to find the pressure-induced superconductivity in type-II WSM candidate TaIrTe$_4$. Our complementary measurements of high-pressure resistance, magnetoresistance, *ac* susceptibility, Hall coefficient and synchrotron X-ray diffraction indicate that the superconductivity emerges at ~23.8 GPa, around which the response of its positive Hall coefficient to pressure turns its tendency from the increase to the decrease and the positive magnetoresistance disappears. Our high-pressure structure studies reveal that at this critical pressure, the lattice distorts apparently along $c$ axis, which leads to the change in topology of band structure and in turn drive the superconducting transition. Whether a Weyl superconducting state exists in the WSM candidate TaIrTe$_4$ is an open question, which deserves further investigations by further experiments and sophisticated theories.

**Methods**

High-quality single crystals of TaIrTe$_4$ crystals were grown using the flux method using Te as the flux[73]. Ta powder, arc-melted Ir and Te chunks were loaded into a 5-mL alumina crucible with the molar ratio of Ta : Ir : Te = 3 : 3 : 94. The crucible was then sealed inside a quartz tube under 1/3 of atm of Ar. The ampule was heated up to 1200 °C, stayed for 3 hours and then cooled to 550 °C at a rate of 2 °C/h.

High pressure resistance measurements below 40 GPa were performed in a diamond anvil cell (DAC), in which diamond anvils with 300 $\mu$m flats and a nonmagnetic rhenium gasket with 100-$\mu$m-diameter hole were adopted. The standard four-probe electrodes were applied on the cleavage plane of the TaIrTe$_4$ single crystals. To provide a quasi-hydrostatic pressure environment for the sample, NaCl powder was employed as the pressure medium. For the higher pressure resistance measurements above 40 GPa, we employed a diamond anvil cell with 200 $\mu$m flats on which the standard four-probe technique was also used. High-pressure Hall coefficient measurements were carried out by the standard method. The sample with a rectangular shape was loaded in a DAC. To keep the sample in a quasi-hydrostatic pressure environment, NaCl powder was employed as the pressure medium. Because the transport properties of TaIrTe$_4$ show highly anisotropy at ambient pressure[62-65], we applied current along the *b* axis in all our resistance and Hall measurements. The high-pressure alternating-current (*ac*) susceptibilities were detected using a primary/secondary-compensated coil system surrounding the sample[44]. High-pressure X-ray diffraction (XRD) measurements were performed at room temperature at beamline 4W2 at the Beijing Synchrotron Radiation Facility and at beamline 15U at the Shanghai Synchrotron Radiation Facility, respectively. Diamonds with low birefringence were selected for these XRD measurements. A monochromatic X-ray beam with a wavelength of 0.6199 Å was employed and silicon oil was taken as a pressure-transmitting medium. The pressure for all measurements below 40 GPa were determined by the ruby fluorescence method[74], and for all measurements above 40 GPa were determined by the shift of diamond Raman[75,76].

**Acknowledgements**

We thank Dr. Li Zhang for helpful discussions on our high-pressure XRD data. The work in China was supported by the National Key Research and Development Program of China (Grant No. 2017YFA0302900, 2016YFA0300300 and 2017YFA0303103), the NSF of China (Grants No. 11427805, No. U1532267, No. 11604376), the Strategic Priority Research Program (B) of the Chinese Academy of Sciences (Grant No. XDB25000000). The work at UCLA was supported by the U.S. Department of Energy (DOE), Office of Science, Office of Basic Energy Sciences under Award Number DE-SC0011978.

†Correspondence and requests for materials should be addressed to L. Sun (llsun@iphy.ac.cn)

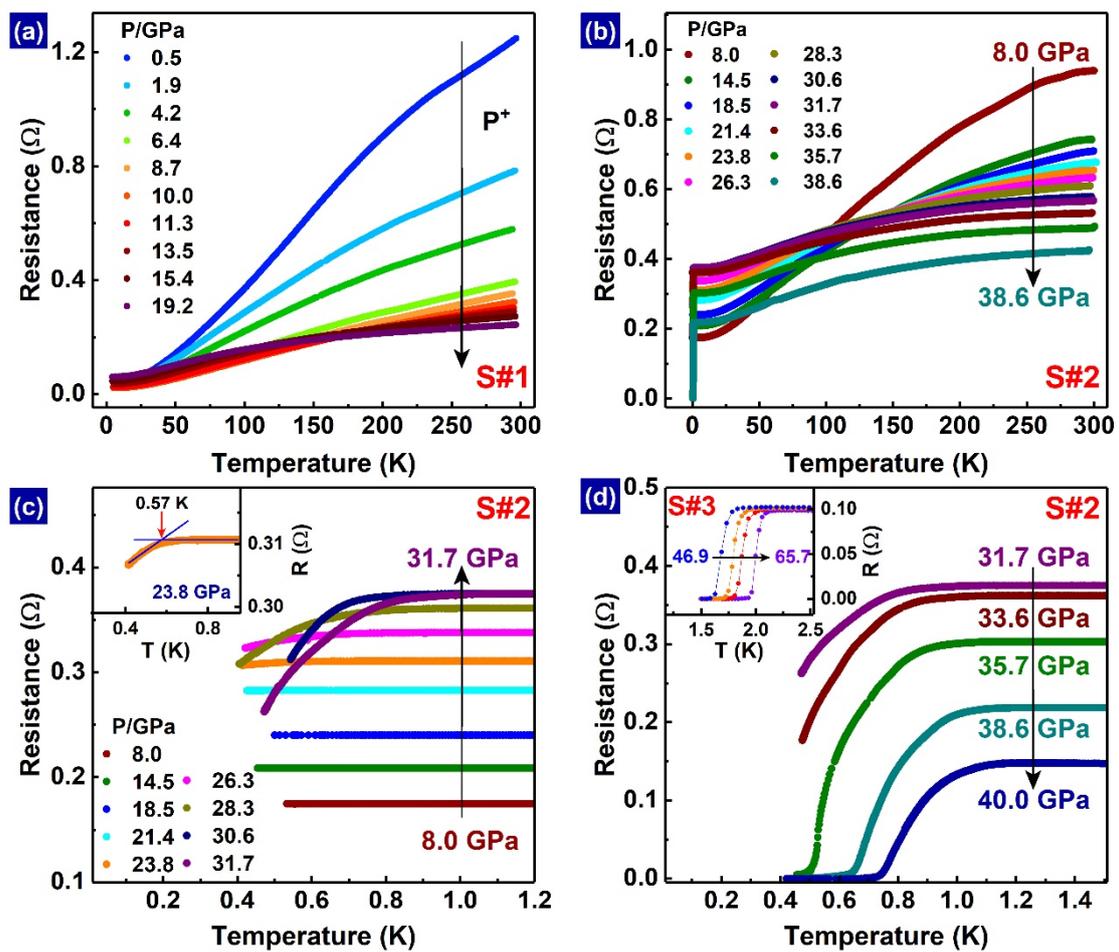

**Figure 1 Temperature dependence of electrical resistance of TaIrTe$_4$ at different pressures.** (a) and (b) display resistance as a function of temperature up to 19.2 GPa for the sample 1 (S#1) and 65.7 GPa for the sample 2 (S#2). (c) and (d) show the low temperature resistance of figure b, displaying the superconducting transitions at higher pressures. The inset of figure d exhibits the superconducting transition observed from sample 3 (S#3).

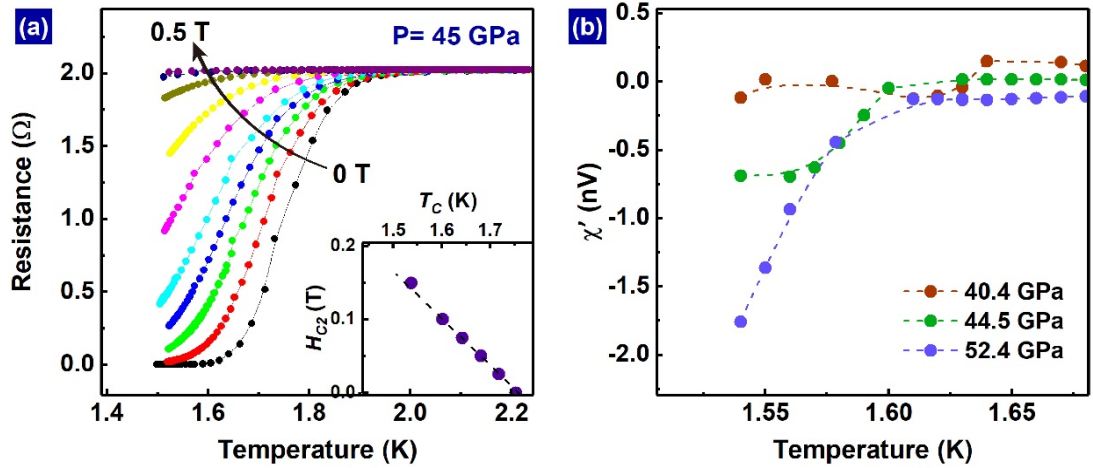

**Figure 2 Characterizations of pressure-induced superconductivity in WSM candidate TaIrTe$_4$.** (a) Magnetic field dependence of superconducting transition temperature measured at 45 GPa. The inset shows upper critical field $H_{c2}$ as a function of superconducting transition temperature $T_C$ for pressurized TaIrTe$_4$. (b) The results of high-pressure *ac* susceptibility measurements.

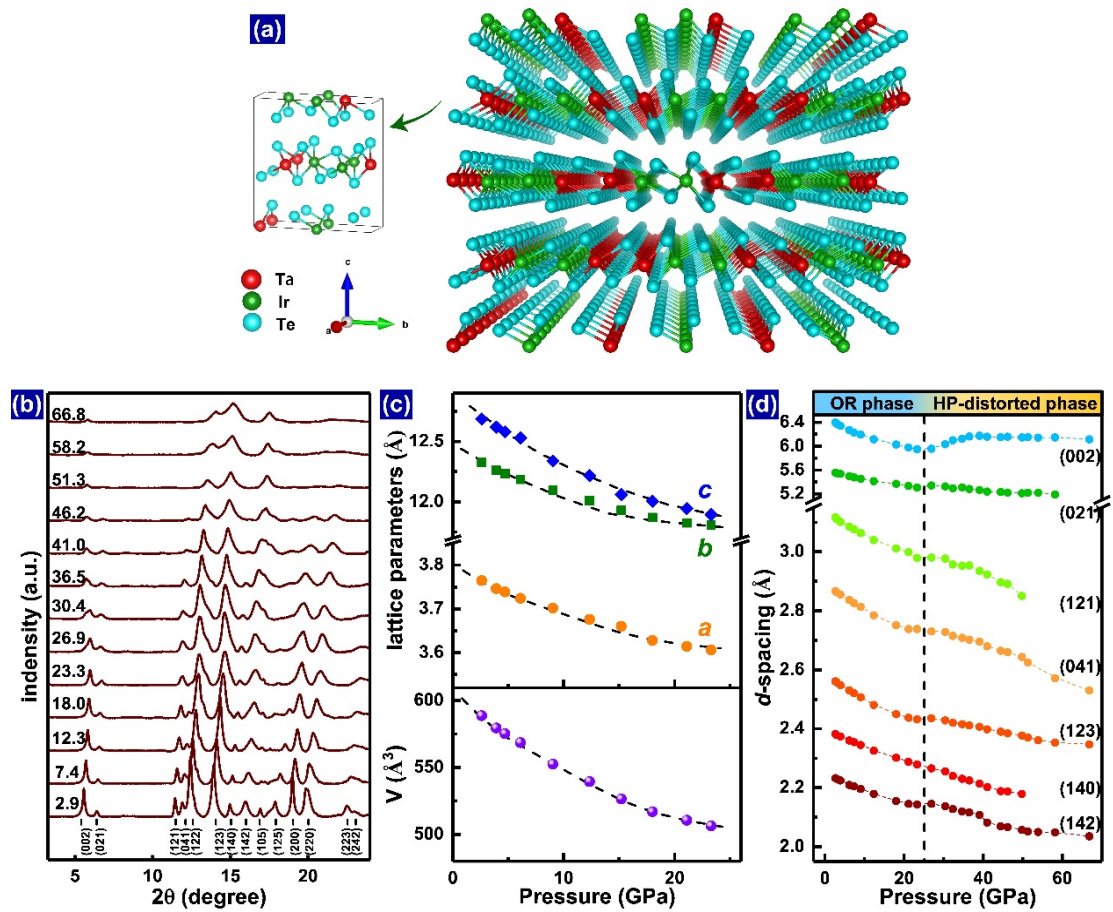

**Figure 3 Structural information for pressurized WSM candidate TaIrTe$_4$.** (a) Schematic crystal structure of TaIrTe$_4$. In the crystallographic description, TaIrTe$_4$ crystallizes in an orthorhombic unit cell. (b) X-ray diffraction patterns collected at different pressures. (c) and (d) Pressure dependence of lattice parameters for the orthorhombic TaIrTe$_4$ up to 26.9 GPa. (e) Plots of *d*-spacing value versus pressure which are extracted from the high-pressure X-ray diffraction data.

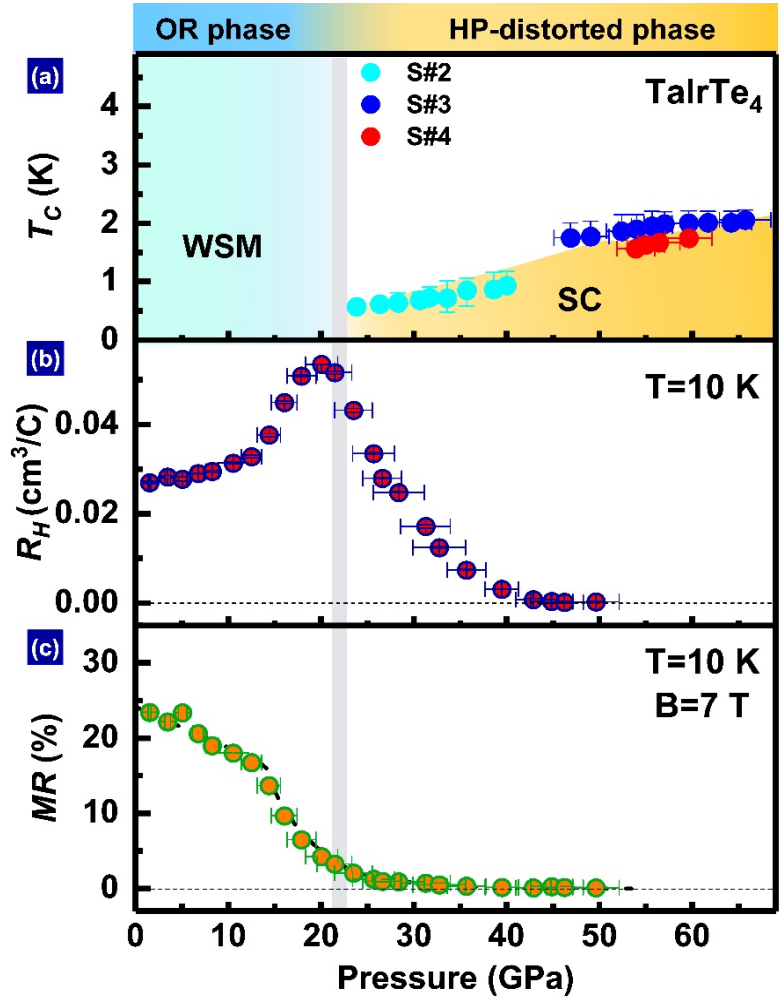

**Figure 4 Summary of experimental results of WSM candidate TaIrTe$_4$.** (a) Pressure -$T_C$ phase diagram with structure information for TaIrTe$_4$. WSM and SC represent Weyl semimetal and superconducting states, respectively. S#2, S#3 and S#4 stand for sample2, sample 3 and sample 4 (see data of S#4 in the SI). (b) Pressure-dependence of Hall coefficient measured at 10 K. (c) Magnetoresistance (*MR*) as a function of pressure measured at 10 K, where *MR%=[R(7T)-R(0)]/R(0T)×100%*.